\documentclass[11pt]{amsart}
\usepackage{geometry}                
\usepackage{graphicx}
\usepackage{amssymb,txfonts}
\usepackage{epstopdf}
\usepackage{caption}
\title{The Puzzle of Empty Bottle  in Quantum Theory}
\author{Bogdan Mielnik}

\begin{document}
\begin{abstract}
We discuss an extremely simple effect of 'shadowing' where the very existence of the measuring 
apparatus deforms the evolution of quantum states even if the measurement is never preformed. 
In spite of strange intuitive aspects, it might be related to some recent doubts about
the completeness of quantum theories.  
 
\end{abstract}
\maketitle

\section{Introduction}

While the quantum paradoxes owe  their  origin  to  the  Schr\"odinger's thought  \cite{Schro} 
  (questioning   the   consistency   between   the evolution and  measurement laws), the question whether the quantum measurements consist in sudden jumps \cite{Sudbery1} or some microobject inststability 
\cite{Haroche}, or just in the decoherence \cite{Coherence} awake less attention. The last decades evolved differently, with interest focused on non-locality problems, starting from the 
historical Einstein--Rosen--Podolski   (EPR) result   \cite{EPR}. Yet, the subsequent discussions \cite{AA1,AA2,2bot,Corrigendum,Fink,Mosley}, illustrate indeed that many unsolved problems  return persistently, 
still without final conclusions. The surprising consequence EPR \cite{EPR} was  the teleportation \cite{Bennet,Gis}, as well as uneasy  problems with   the   Wheeler's paradox of 
"delayed choice experiment" at the bottom \cite{Wheeler}. No less unexpected  effecs were described   by  Elitzur and  Vaidman  \cite{EV,Vaidman,VaidmanB},  permitting  to  "see   in   the   dark" \cite{Kwiat}. 

  Worse, since  
an evidence is also accumulating  that the formalism of quantum theories is not so universal as the several generations believed.

\section{Linear navigation?}

The fact which decided  the form of the present quantum theories was the observation of 
the interference patterns of  particle beams suggesting the picture of the linearly propagating 
de Broglie waves which after diffracting on the material obstacles paint the interference 
fringes (but see the discussions between de Broglie and Einstein \cite{Norsen})  The subsequent generalization leads to an abstract scheme of 
the wave-particle duality in which the (pure) microparticle states are represented by vectors 
in  linear {\em Hilbert} spaces, obeying the {\em superposition principle}, evolving always  
according to a certain linear law, the process which can be interrumpted by the statistical 
measurements, with probabilities  defined     
by the quadratic forms of the state vectors.   

A persistent desire  to deal with the linear propagation (except of  measurements)  
determined also the description of multiparticle states. In fact, any 
equation  can be linearized at the cost of multiplying the number of variables. In case of 
many particles this was achieved by introducing the {\em entangled states} represented 
by vectors of the tensor product spaces, in which the basic propagation was again linear and   
probabilities given again by the quadratic forms, or alternatively, by the {\em observables} 
 represented by self-adjoint operators.  The resulting picture, modulo  
postponed interpretational problems, for a long time seemed  
a universal form of quantum theories. The deeper troubles were always implicit  in 
 Wheeler's  "delayed choice experiments", 
tolerated usually  as a pintoresque (or even surrealistic) anecdote,   
far away from the typical physicist work. Yet, the 
difficulty is not smaller in the  {\em interaction free measurement \cite{EV}} 
invading as deeply our intuitions. Here,  it seems essential that the final 
{\em to be or not to be} effect happens for each single photon. However, what is 
the {\em single photon}?

To give to the dilema even more extreme character, we permit urselves to 
imagine the same EV experiment with a pair of parallel fibers quite long (see Fig.1. I beg the reader 
to forgive me this element of S/F story; we all know that there are no interstellar fibers!) 
However, assume, two of them are quite long. The next, (much shorter) part of the trajectory leads 
to the second beam splitter. An alternative trajectory has also its longer part: they both meet in the 
second splitter, opening the way to either the detector $\bf D_1$ or $\bf D_2$.  Now, if there is no obstacle in one of the trajectories, then the photon state would split into two 
coherent, but weaker  components which would finally join at the second splitter (happily) 
reconstructing the initial photon state, falling into $\bf D_1$.  
   
All this is rather easy to imagine if the photon is a very short pulse. However, can the single photon propagate just as a pulse? Or perhaps, one shold it imagine rather as a very long, narrow wave 
divided by the first splitter into a pair of weaker but very long components which 
laboriously and slowly reconstruct their initial form at the second splitter, falling then (slowly) to the 
detector  $\bf D_1$? If so, the problem would arise, at which moment the detector responds 
to our {\em single photon} ? At the beginning or at the end of the process? 

Worse, because if one of the (EV) trajectories is blocked by the bomb, then after what time the bomb 
explodes, but if not, then after what time the (long but incomplete) photon trajectory which {\em would 
cross the bomb} is misteriously annihilated and contributes (again misteriously) , to the other weak 
component creating the propagating (complete) 1 photon state, whicch however arrive to 
another detector? We can only conclude that  the story is incomplete: indeed, it is impossible to 
form any mental picture of the sacred {\em linear propagation} of the photon before the 
experiment is finished. And when finished, it  can give an information about an obstacle which exists precisely in the place where the photon newer was! Here let me remind the point 
made by Sudbery  \cite{Sudbery3}: 
\bigskip

\begin{center}
\begin{minipage}[h]{0.6\textwidth}
"It is often stated that however puzzling some of its 
    features may be, quantum mechanics does  constitute  a 
    well  defined  algorithm  for   calculating   physical 
    quantities. Unless some form of continuous  projection 
    postulate is included as a part of the algorithm, this 
    is not true."  
\end{minipage}
\end{center}

\bigskip

A question of course remains: what is the photon wave function in the optical fiber? can it be a kind of extenive creature, many kilometers long, crawling along the fiber? It turns out that the description of the 
photon waves in the optical fibers is already known from the paper of I. Bialynicki-Birula \cite{IBB} 
described not by any plane wave, but by a Bessel function cf. eq (55) (which seems a significant progress comparing to the questionable visions of Quarks as the plane waves running inside of the nucleon surfaces!)    

However, except of the propagation in single fibers this does not sole the mystery of the propagation of the photon waves in \cite{EV} and indeed, it cannot by any local theory, although the final result 
gives an important information (about the obstacle). A sequence of studies on the imperfect 
cases of EV bombis recently undertaken \cite{Seth,Sci1,Sci2,QF}, and at least one suggests that the experimants with linearly propagating entangled states must affect the past.

 In what follows, 
our aim is  to forget for a moment about the locality problems, returning to the 
"historical roots", where  some consequences  of the Schr\"odinger's 
paradox still wait to be explored. 

\begin{figure}[h]
\includegraphics[scale=0.5]{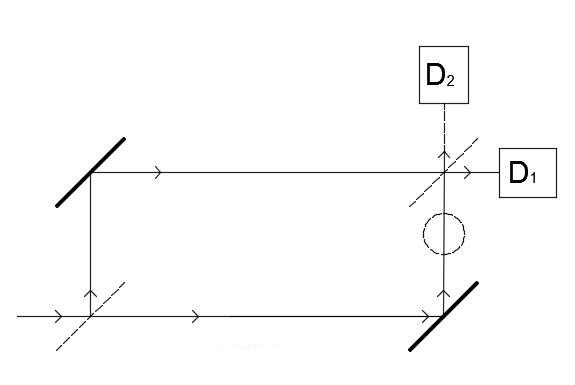}
\captionsetup{labelformat=empty}
\caption{Fig 1}
\end{figure}

\section{Half full, half empty.}

Our story concerns a quantum system in a superposed energy state,  which will be 
reduced -- though not when  the  experimentalist  decides, but when the system itself 
decides by emitting a photon (compare with the 'time of arrival' \cite{Screen,Leavens1,Leavens2}). As a 
simplified model, we consider a bottle containing an atom in a state  {\em superposed} 
 of two lowest energy eigenstates, ground state \(\phi_0\) and  an excited 
state \(\phi_1\). If the atom is in the excited state, we shall  say  that  "the bottle is full",  
but if in the ground state, "the bottle is  empty".  

In some distant past, the experimentalists examining the spectral lines imagined   
an atom always in one  of the energy  eigenstates. Today the picture is different.  The existence of the superposed 
(but pure) energy states is  unavoidable  if one takes seriously the quantum mechanical 
formalism. The bottle  is  only  to assure that the atom is left in peace. The only thing it can  do  is  to radiate, settling itself on the ground state \(\phi_0\). We shall  assume,  that  at some  safe  distance,  there  is  a  sensitive screen in the bottle, prepared to detect the photon, should the atom radiate. 

\begin{figure}[h]
\includegraphics[scale=0.3]{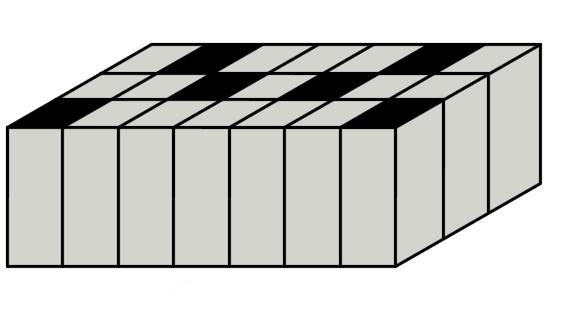}
\captionsetup{labelformat=empty}
\caption{Fig 2}
\end{figure}

In almost all ensays on the atom radiation one can find  the description of the excited states as some  narrow superpositions of slightly different energy eigenstates, forming an unstable state, 
whose average lifetime $\tau$ is inverse to the (little) energy width $\Delta E$, 
in agreement with the time-energy uncertainty, even though, the last principle awoke a lot of 
unfnished discussions \cite{T1}. Anyhow, by reading books on the excited state decay one can always find 
the considerations in which the beginning of the decay process is an excited state  
with a very narrow  spectral line. However, I have never seen a study of a decay process  
starting from the superposition  of two very distant energy levels. One might think that the difference is 
superflous, but the excited state decay might not fulfill  the principle of the linear drift, and anyhow  
why anybody considered the {\em coherent superposition} of two distant  levels as the  starting point of the decay process?

To fix  attention, let us assume  that  our  initial  state  \(\phi\) is  an equitative superposition \(\phi =  a_0\phi_0  +  a_1\phi_1\),  where  \(\vert a_0\vert^2 =  \vert a_1\vert^2=  1/2\) (bottle half full, half empty). From a credible  phenomenology  we  know the behaviour of an atom in its ground state  \(\phi_0\).  (If  unperturbed,  it just remains in \(\phi_0\) forever, \(\phi_0(t) = exp(-itE_0)\phi_0\) ). We  know  as  well  the behaviour of the excited state \(\phi_1\). On the level of purely  quantum mechanical description, this state is as stationary 
\begin{equation}
                      \phi_1(t)=exp(-itE_1)\phi_1                  
\end{equation}
In reality though, the stationary evolution might be interrupted  by  an unpredictable photon emission with a sudden jump to the ground state $\phi_0$. For an ensemble of pure excited  atoms,  the  number \( N(t) \) of  the  ones surviving in the initial state $\phi_1$ will be  decreasing  exponentially  in time, \(N(t) = exp(-\gamma t)N(0)\), where \(1/\gamma\) is the average life time of $\phi_1$; the survivors   landing gradually in the ground state \(\phi_0\). However, what  happens in a process with atoms  in an initial superposed energy  state  \(\phi  =  a_0\phi_0  +  a_1\phi_1\) (the bottle {\em neither full, nor empty})? At the first sight,  it  may  seem that there is hardly any problem here: the evolution  of  the  atom must simply obey the standard law 
\begin{equation}
               \phi(t) = a_0 e^{-iE_0t}\phi_0 + a_1 e^{-iE_1t}\phi_1
\end{equation}
granted by the superposition principle,  except  if  it  shall  suddenly radiate emitting a photon of energy $E_1-E_0$, colocating itself on  the  ground  state \(\phi_0\).  We  shall  show, however, that this plausible picture contains certain mysteries. 

Suppose that in some atoms suddenly happens a spontaneous ({\em introspective}?) state reduction. They find themselves in the excited state \(\phi_1\) and hence, can emit the photon of the energy $E_1-E_0$. However, average energy of the superposed initial state is less than $E_1-E_0$. 
The question arises,  whether they must ask some energy credit from their detector? If so, is it some influence 
of the detector due to its very existence, even if the measurement is not performed \cite{Gisin2,Gis}, or   a kind of {\em shadoving} \cite{Jadczyk}?. Of course, not all {\em semi-excited} atoms can radiate. 
But even if the total energy balance is not violated, the single  atom behavior has still some mystery.

\section{The effect of vanishing hope...}

It may be amazing to imagine a population of \(N \) atoms in the initial state \(\phi = a_0\phi_0 + a_1 \phi_1\).  Assume now  a  gedanken scenario, in which every atom is closed in its own bottle, in form of  a little, mesoscopic cell. Assume moreover, that the top surface of the cell is simultaneously a detector, sensitive to  the  photons  of the particular energy \(\hbar \omega = E_1-E_0\). If an atom radiates,  the  top  of its  cell  turns  black  (it  is  burned!).  By  calculating  the (increasing) number  of the black cells, we know how many atoms have already radiated (Fig.2). 
If all atoms are initially in an identical  superposed state \(\phi= a_0\phi_0 + a_1\phi_1\), then if somebody performed a check at the very beginning, he would  find  50\% of them in the ground state \(\phi_0\), and henceforth, unable to radiate. However, if no initial test was performed at \(t=0\), then anyhow  50\%  of  the  atoms  will  never radiate.  Thus, for $t \rightarrow +\infty $ all atoms must end up in the ground state \(\phi_0\), though for different reasons: 1/2 of them, since they  have  radiated  and  settled down in \(\phi_0\); the  remaining  1/2,  just  because  the  long  waiting  was equivalent to a yes-no measurement (i.e., to ask whether  the  atom  was able at all to radiate), the negative answer reducing the  \(\phi_1\)  component to nonexistence, even though no photon was emitted.  As  the  matter  of fact, what has caused the state collapse in this last case, was not  any active external intervence, but just the vanishing hope ( that  the  atom could have been in the excited state  \(\phi_1\)).  Indeed,  supposing  that  the average lifetime of the atom in the excited state \(\phi_1\), e.g., is \(1min\), but the atom in the initial state \(\phi\)did not radiate for 100 years, then,  we can be  certain that it will radiate  never.  According  to  quite orthodox statistical interpretation, this certainty means that the  atom state can no longer correspond to \(\phi\), but it must be practically  identical  to \(\phi_0\). 
\bigskip

Even if the global energy balance is not affected, and even if we disregarded the principle of not 
postselecting,  the situation seems  extremely strange. While the atoms which have radiated cause almost no problem, the ones which did'nt contain a puzzle!  Their  superposed energy state  vanished, giving place to \(\phi_0\). The  only  external  factor was our vanishing hope (take it as a rhetoric figure  if  you  dislike!), or perhaps shadowing \cite{Jadczyk}? Anyhow, the bottle was {\em half full, half empty }, nothing escaped,  yet  the bottle is {\em empty!} 
Note that all difficulties would vanish if we simply assumed that no coherent superposition of two 
distant bound states can exist (remember Einstein boxes \cite{Norsen}?)
Yet it would be too risky to derive from these alegoric remarks too premature conclusions. Anyhow, they 
seem to show that our theories contain some elements which are only our mental constructs: why they sometimes help (de Broglie waves!) and sometimes not, we still ignore.


\end{document}